\documentclass{aa}  

\usepackage{graphicx}
\usepackage{txfonts}
\usepackage{natbib} 
\bibpunct{(}{)}{;}{a}{}{,} 

\usepackage{verbatim} 
\usepackage{hyperref}   
\hypersetup{colorlinks=true,linkcolor=blue,citecolor=blue,filecolor=blue,urlcolor=blue}

\begin{document}

   \title{Tests and calibrations  of stellar models with two triply eclipsing triple systems} 

   \subtitle{}
   \author{G. Valle \inst{1, 2}\orcid{0000-0003-3010-5252}, M. Dell'Omodarme \inst{1}\orcid{0000-0001-6317-7872}, P.G. Prada Moroni
        \inst{1,2}\orcid{0000-0001-9712-9916}, S. Degl'Innocenti \inst{1,2}\orcid{0000-0001-9666-6066} 
}
\titlerunning{Analysis of two triply eclipsing triple systems}
\authorrunning{Valle, G. et al.}

\institute{
        Dipartimento di Fisica "Enrico Fermi'',
        Universit\`a di Pisa, Largo Pontecorvo 3, I-56127, Pisa, Italy\\
        \email{valle@df.unipi.it}
        \and
        INFN,
        Sezione di Pisa, Largo Pontecorvo 3, I-56127, Pisa, Italy
}

   \date{Received 04/10/2024; accepted 04/02/2025}

  \abstract
{}
{
We investigated the possibility of using two recently characterised triply eclipsing triple systems to constrain stellar model parameters. We specifically focused on evaluating the influence of the underlying astrophysical assumptions employed in the characterisation of the system to fix absolute values of the radii, effective temperatures, and metallicity. }
{We used dense grids of pre-computed stellar models to
fit the data for the triply eclipsing systems with a modified version of the SCEPtER pipeline }
{We achieve an excellent agreement with observational data for TIC 650024463, which comprises three low-mass main-sequence (MS) stars. We find it has an age of $9.0^{+1.4}_{-1.1}$ Gyr and a multimodal posterior density. 
Characterising TIC 323486857 proved more challenging. This system comprises two intermediate-mass MS stars and a slightly more massive tertiary in the red giant branch phase. For this last system we tested alternative scenarios for convective core overshooting. When all stars were assumed to have the same overshooting efficiency, significant discrepancies arose with the observed data for the tertiary star. This discrepancy may arise from the different assumptions regarding overshooting  efficiency made for the observational characterisation of the system, in which an increasing overshooting efficiency with stellar mass was adopted. By allowing independent overshooting efficiencies for all stars, we recovered a solution close to that adopted in the system observational characterisation.
Encouragingly, despite the relevant differences between the adopted stellar models and those used for the observational characterisation, we found a system age of $2.33^{+0.18}_{-0.16}$ Gyr  in all the tested scenarios, and this age is in agreement with independent determinations.
}
{} 
   \keywords{
Binaries: eclipsing --
Stars: fundamental parameters --
methods: statistical --
stars: evolution --
stars: interiors
}

   \maketitle

\section{Introduction}\label{sec:intro}

Triple-star systems represent the simplest multi-star configurations beyond binaries, offering a valuable opportunity to compare stellar model results with observational data.
When the third body in a triple system produces eclipses, significant information can be gleaned. These events occur when the inner eclipsing binary occults the third star or vice versa. By analysing the eclipsing binary light curves, extracted measurements of eclipse timing variations, and energy distribution data from archival surveys, it is often possible to determine key stellar parameters, including masses, radii, and effective temperatures, along with orbital parameters. This can be achieved with limited reliance on extensive radial velocity (RV) measurements \citep[see e.g.][]{Carter2011, Alonso2015, Rappaport2023, Rappaport2024}. Precision of a few percent in mass and radius determinations is often possible for these systems. 

While the precision of these parameters may be lower than that achievable for eclipsing binaries through extensive spectroscopic analysis, triple systems are still a valuable test bed for stellar models. Testing and calibrating stellar models with binary systems often demands exceptional observational precision, typically below 1\% \citep[see e.g.][]{TZFor, Miller2020, Helminiak2021, Anders2023}. The additional constraints provided by a third star in triple systems can aid in refining estimates of stellar evolutionary parameters, even with lower precisions in mass and radius determinations.
Studies of triply eclipsing triple systems \citep[e.g.][]{Carter2011, Feiden2011, Derekas2011} exemplify the valuable role a third star can play in refining system solutions.

The census of triply eclipsing triple systems has steadily grown in the last decade. \citet{Rappaport2024} present observations of seven triply eclipsing triple systems. Their analysis leveraged Transiting Exoplanet Survey Satellite data along with archival data from different sources. System parameters were extracted via a photodynamical analysis using \textsc{Lightcurvefactory} \citep{Borkovits2019}.  No spectroscopically determined temperatures
or metallicities were available for the systems; therefore, absolute temperatures and radii were estimated  in part using spectral energy distribution fitting. The parameters are, however, astrophysical model dependent because PARSEC isochrones \citep{Bressan2012} were used as part of input information for the fit to determine the absolute radii and metallicities of the systems.  While the mass, radius, and effective temperature determinations were obtained without RV measurements, it has been suggested  \citep{Borkovits2022, Rappaport2024}  that the available excellent eclipse timing variation measurements make their estimation  extremely robust.

Investigating these triply eclipsing systems is interesting for testing the reliability of this claim under different scenarios. 
Stellar models are expected to show the best agreement for low-mass main-sequence (MS) stars that lack a convective core \citep{incertezze1, Stancliffe2015}. 
The underlying assumptions within the models have a minimal impact in these cases. However, for more evolved and massive stars, the treatment of convective core overshooting significantly affects their evolution. Consequently, the choice of the astrophysical model becomes increasingly important in determining the system's parameters for these stars.

Among the \citet{Rappaport2024} systems, two offer a good opportunity to test stellar models accuracy and to check the impact on the fit of the convective core overshooting parameter. A first system, TIC 650024463, is composed of three MS stars with clearly different masses in the range [0.68, 0.88] $M_{\sun}$. 
The system is nearly flat and characterised by an inner binary period of about 7.2 d (semi-major axis of the orbit of 18 $R_{\sun}$) and an outer period of about 109 d (semi-major axis of the tertiary orbit of 127 $R_{\sun}$). The outer eccentricity is $e \approx 0.33$.
This system provides a good opportunity to check the reliability of simultaneous stellar model predictions for stars without a convective core. 

The second system,  TIC 323486857, hosts more massive stars in the range [1.2, 1.6] $M_{\sun}$; two are MS stars, and the third, the most massive, is on the red giant branch (RGB). 
The system has a short outer orbital
        period of 41 d, a low outer eccentricity $e = 0.0066$,
        and is  nearly flat. The period of the inner binary is  0.88 d, with a semi-major axis of 5.4 $R_{\sun}$, while the semi-major axis of the tertiary orbit is about 81  $R_{\sun}$.
This system is a good target to investigate how the underlying astrophysical assumptions on the  
convective core overshooting efficiency influences the system fit when  different stellar models are adopted. 
While a more comprehensive analysis of a larger sample of triply eclipsing triple systems would be ideal,
the computational burden imposed by the fit of these system, partly due to the observational uncertainties being significantly greater than those for double-lined eclipsing binary systems, is so high that we were forced to restrict the investigation to these two specific targets. 
The dimensionality of the stellar model grid required to accurately fit a system is primarily determined by the observational precision of the component masses. For well-characterised double-lined eclipsing binaries with extensive RV measurements, these uncertainties are often of the order of 0.001 $M_{\sun}$, allowing the stellar mass to be effectively fixed in the grid computation. This is not the case when observational errors exceed 0.01 $M_{\sun}$, as a broader mass range encompassing approximately 3 $\sigma$ around the observational constraint must be covered, necessitating a significantly larger grid.

The structure of the paper is as follows. In Sect.~\ref{sec:method} we present the grid of stellar models and the fitting method used in the estimation process. The results for both systems are presented in Sect.~\ref{sec:results}.
Concluding remarks can be found in Sect.~\ref{sec:conclusions}.

\section{Methods}\label{sec:method}
\subsection{Stellar model grid}

The grid of stellar evolutionary models was calculated for the mass range 0.60 to 1.00 $M_{\sun}$ for the first system and 1.10 to 1.70 $M_{\sun}$ for the second one, spanning the evolutionary stages from the pre-MS to the RGB tip.
The initial metallicity [Fe/H] was varied from $-0.4$ dex to 0.4 dex with
a step of 0.02 dex. 
We adopted the solar heavy-element mixture by \citet{AGSS09}. 
For each metallicity, we considered a range of initial helium abundances based on the commonly used linear relation $Y = Y_p+\frac{\Delta Y}{\Delta Z} Z$
with the primordial helium abundance  $Y_p = 0.2471$ from \citet{Planck2020}.
The helium-to-metal enrichment ratio $\Delta Y/\Delta Z$ was varied
from 1.0 to 3.0 with a step of 0.25. 

The models were computed with the FRANEC code, in the same
configuration as was adopted to compute the Pisa Stellar
Evolution Data Base\footnote{\url{http://astro.df.unipi.it/stellar-models/}} 
for low-mass stars \citep{database2012}. The only difference with respect to those models is that  the outer boundary conditions  were set by the \citet{Vernazza1981} solar semi-empirical $T(\tau)$, which approximate well results obtained using the hydro-calibrated $T(\tau)$ \citep{Salaris2015, Salaris2018}.
The models were calculated assuming the solar-scaled mixing-length parameter $\alpha_{\rm ml} = 2.02$.
The extension of the extra-mixing region beyond the Schwarzschild border
was parametrised  in terms of the pressure scale height $H_{\rm 
        p}$: $l_{\rm ov} = \beta H_{\rm p}$, with 
$\beta$ from 0.00 to 0.28 with a step of 0.01. The code adopts step overshooting assuming an instantaneous mixing in the overshooting treatment. The radiative temperature gradient is adopted in the overshooting  region \citep[see][for more details of the overshooting implementation]{scilla2008}.   
        Atomic diffusion was included adopting the coefficients given by
\citet{thoul94} for gravitational settling and thermal diffusion. 
To prevent extreme variations in the surface chemical abundances, 
the diffusion velocities were
multiplied by a suppression parabolic factor that takes a value of 1 for 99\% of the mass of the structure and 0 at the base of the atmosphere \citep{Chaboyer2001}.

Raw stellar evolutionary tracks were reduced to a set of tracks with the same number of homologous points according to the evolutionary phase.
Details about the reduction procedure are reported in the appendix of \citet{incertezze1}.  Overall, to achieve the required grid density for all the investigated parameters,  more than $10^5$ stellar tracks were computed. 

\subsection{Fitting technique}\label{sec:fit-method}

The analysis was performed adopting a modified SCEPtER pipeline\footnote{Publicly available on CRAN: \url{http://CRAN.R-project.org/package=SCEPtER}, \url{http://CRAN.R-project.org/package=SCEPtERbinary}}, a well-tested technique for fitting single and binary systems \citep[e.g.][]{TZFor, Valle2023a}. 
The pipeline estimates  the parameters of interest (i.e. the system age,  its initial chemical abundances, the convective core overshooting parameter) adopting a grid maximum likelihood  approach.

The method we used is explained in detail in \citet{binary}; here, we discuss only the modifications needed to fit a triple system. For every $j$-th point in the fitting grid of pre-computed stellar models, a likelihood estimate is obtained for all stars,
\begin{equation}
        {{\cal L}^{1,2,3}}_j = \left( \prod_{i=1}^n \frac{1}{\sqrt{2 \pi}
                \sigma_i} \right) 
        \times \exp \left( -\frac{\chi^2}{2} \right)
        \label{eq:lik}
        ,\end{equation}
\begin{equation}
        \chi^2 = \sum_{i=1}^n \left( \frac{o_i -
                g_i^j}{\sigma_i} \right)^2
        \label{eq:chi2},
\end{equation}
where $o_i$ are the $n$ system constraints (masses, radii, effective temperatures and metallicities for the three stars), $g_i^j$ are the $j$-th grid point corresponding values, and $\sigma_i$ are the respective uncertainties. 
The adoption of the $\chi^2$ statistic in Eq.~(\ref{eq:chi2}) to judge the goodness-of-fit is widely used. However, careful consideration is necessary when evaluating the associated degrees of freedom (dofs). In our case, each star is characterised by four parameters, but these are not independent due to the constraints imposed by stellar evolution tracks. \citet{goodness2021} demonstrate that when parameters are linked by an isochrone (and similarly by a stellar track), one dof is lost for each object. This, however, assumes independent observations for each star, which is not the case for multiple systems.
Therefore, accurately determining the actual number of dofs is challenging. In the following, we assume three dofs for each star, and account for a loss of one dof for each estimated parameter, such as initial chemical composition, age, and
$\beta$. 

The joint likelihood of the system is then computed as the product of the three single star likelihood functions.  
The  whole system fit is obtained by imposing that three objects must have a common age (with a tolerance of 1 Myr), identical initial helium abundance, and initial metallicity.
This step gives a pool of models with corresponding likelihood values ${\cal L}_p$. For these models the  temperature ($q_{T;1,2}$ and $q_{T;1,3}$) and masses ratio ($q_{1,2}$ and $q_{1,3}$) are computed. The corresponding fitting values  ${\tilde q}_{T;1,2}$, ${\tilde q}_{T;1,3}$, ${\tilde q}_{1,2}$, and ${\tilde q}_{1,3}$ are accounted for as follows:
\begin{equation}
          {\cal L} = {\cal L}_p  \exp \left( -\frac{\tau^2}{2} \right)
\end{equation} 
\begin{eqnarray}
        \tau^2 &=& \left( \frac{q_{T;1,2} - {\tilde q}_{T;1,2}}{\sigma_{T;1,2}} \right)^2  + 
        \left( \frac{q_{T;1,3} - {\tilde q}_{T;1,3}}{\sigma_{T;1,3}} \right)^2 +
        \left( \frac{q_{1,2} - {\tilde q}_{1,2}}{\sigma_{1,2}} \right)^2+ \nonumber \\ 
        &+& \left( \frac{q_{1,3} - {\tilde q}_{1,3}}{\sigma_{1,3}} \right)^2
.\end{eqnarray}
Since accounting for the radii ratios did not alter the solutions of the studied systems, we excluded it from the final algorithm.
Stellar parameters for models with a likelihood $\cal L$ exceeding 95\% of the maximum likelihood value\footnote{The specific choice of the cut-off value  has minimal impact. Using a weighted mean of the parameters with likelihood as the weight would yield negligible differences.} are averaged and returned as the solution. The error in the fit parameter was evaluated by means of a Monte Carlo procedure, by simulating 5,000 artificial triple systems, These systems were obtained by  Gaussian perturbations of the observational values, taking the correlation between temperatures, metallicities, and masses into account.

\section{Results}\label{sec:results}

\subsection{TIC 650024463}

\begin{table}[ht]
        \centering
        \caption{Constraints for the TIC 650024463 system.} \label{tab:obs-triple1}
        \begin{tabular}{lccc}
                \hline\hline
$T_{\rm eff}$ (K) & $5682\pm100$        & $4608\pm 100$ & $5114\pm100$ \\
$M ~ (M_{\sun})$ & $0.877\pm 0.015$     & $ 0.684 \pm 0.011$ &$0.765 \pm 0.013$\\
$R ~ (R_{\sun})$ & $0.967 \pm 0.015$    & $0.673 \pm 0.007$     & $0.754 \pm 0.009$ \\
$\rm [Fe/H]$ & $-0.1 \pm 0.1$ &&\\
\hline
\end{tabular}
\end{table}             

\noindent Constraints for the TIC 650024463 system, derived from \citet{Rappaport2024}, are presented in Table~\ref{tab:obs-triple1}. In the absence of a spectroscopic determination, errors in effective temperature and metallicity, [Fe/H], are increased to 100 K and 0.1 dex, respectively. Errors in the masses and radii were obtained by computing the geometric means of the  upper and lower error bars given by \citet{Rappaport2024}.

\begin{table}[ht]
        \centering
                \caption{Fit of the TIC 650024463 system.} \label{tab:triple1}
        \begin{tabular}{lcc}
                \hline\hline
                &  S1 & S2 \\
                \hline \\ [-1.8ex]
                $Y$ & $0.260_{-0.003}^{+0.007}$ & $0.283_{-0.009}^{0.007}$ \\[2pt] 
                $Z$ & $0.0109_{-0.0014}^{+0.0016}$ & $0.0128_{-0.0014}^{+0.0021}$ \\ [2pt]
                Age (Gyr) & $9.35_{-1.18}^{+1.42}$ & $8.76_{-1.07}^{+1.26}$ \\ [2pt]
                \hline
                \multicolumn{3}{c}{Best fit parameters}\\
                \hline
                $T_{\rm eff,1}$ (K)& 5667(70)& 5670(65)  \\ 
                ${\rm [Fe/H]}_1$ & -0.19(6) &  -0.10(7)   \\ 
                $M_1 ~ (M_{\sun})$ & 0.88(1) &   0.88(1)   \\ 
                $R_1 ~ (R_{\sun})$ & 0.964(16) &   0.965(15)  \\ 
                $T_{\rm eff,2}$ (K) & 4614(71) &  4630(62)  \\ 
                ${\rm [Fe/H]}_2$ & -0.15(6) &   -0.06(7)   \\ 
                $M_2 ~ (M_{\sun})$ & 0.68(1) &   0.68(1)   \\ 
                $R_2 ~ (R_{\sun})$ & 0.674(5) &   0.673(4)   \\ 
                $T_{\rm eff,3}$ (K) & 5079(78) &   5092(71)   \\ 
                ${\rm [Fe/H]}_3$ & -0.17(6) &   -0.08(7)   \\ 
                $M_3 ~ (M_{\sun})$ & 0.76(1) &  0.76(1)   \\ 
                $R_3 ~ (R_{\sun})$ & 0.755(7) &   0.756(7)   \\ 
                \hline
                $\chi^2$ & 1.65 &   0.78   \\ 
                \hline
        \end{tabular}
\end{table}

\begin{figure*}
        \centering
        \includegraphics[height=5cm,angle=0]{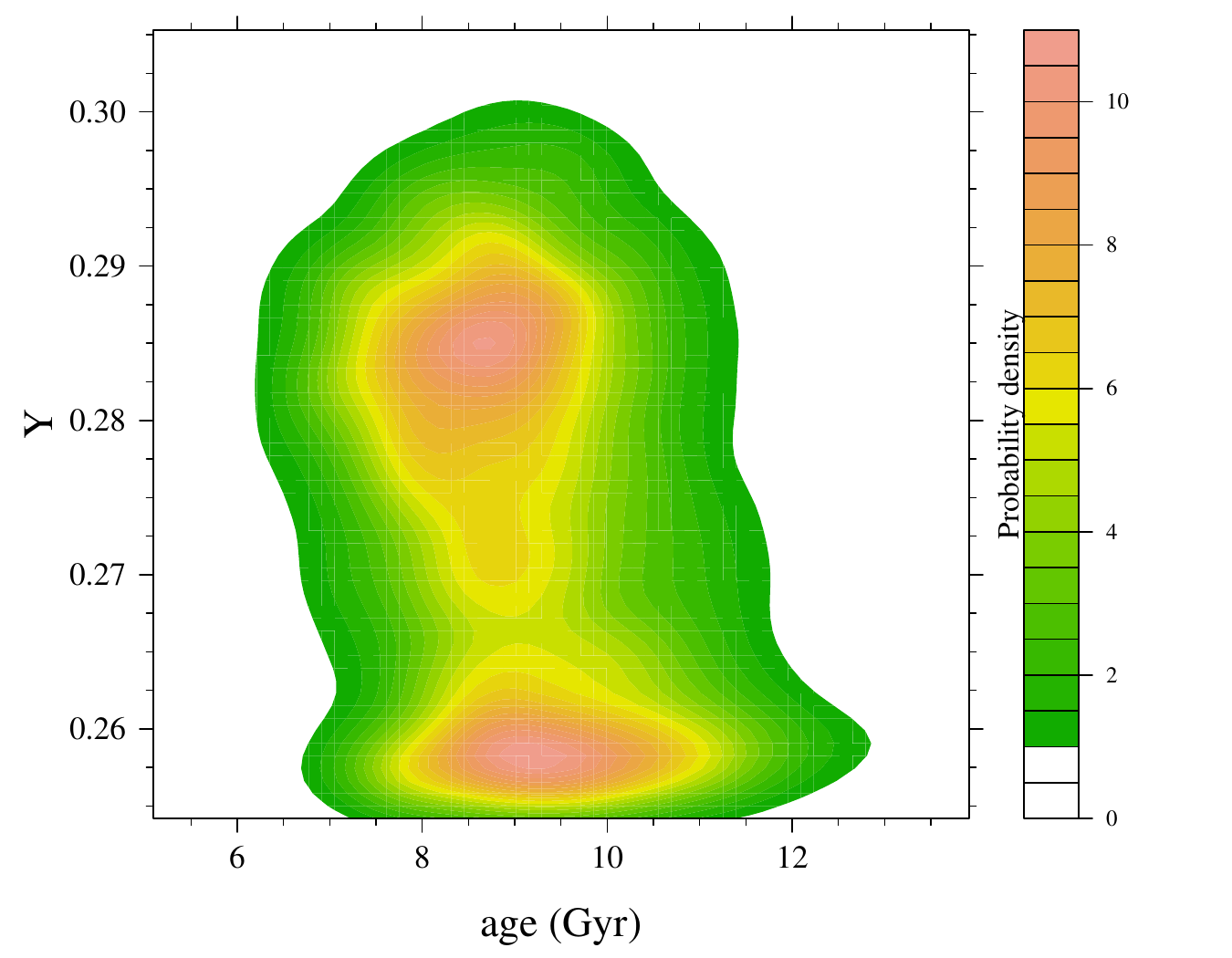}
         \includegraphics[width=5.5cm,angle=0]{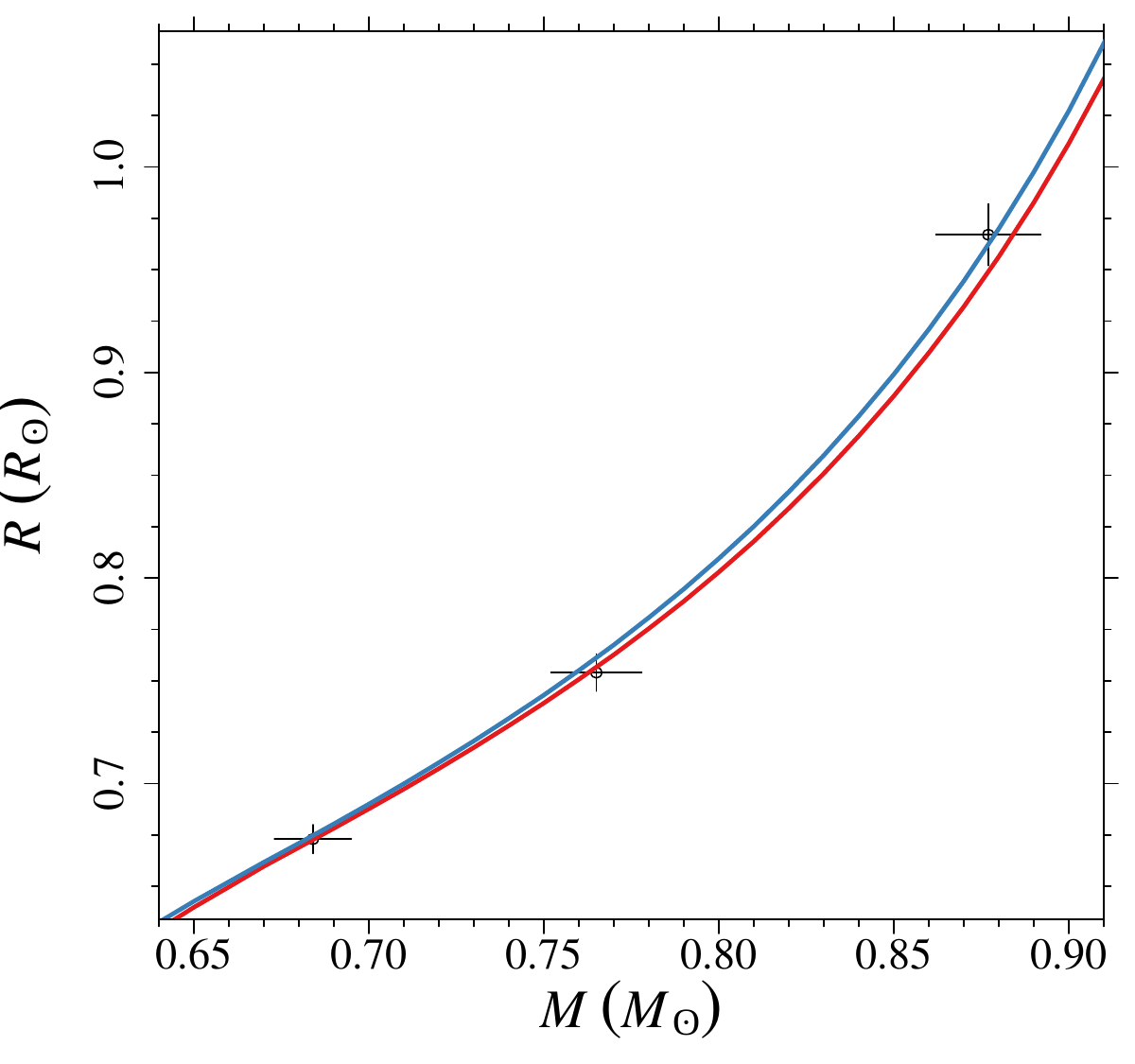}
          \includegraphics[width=5.5cm,angle=0]{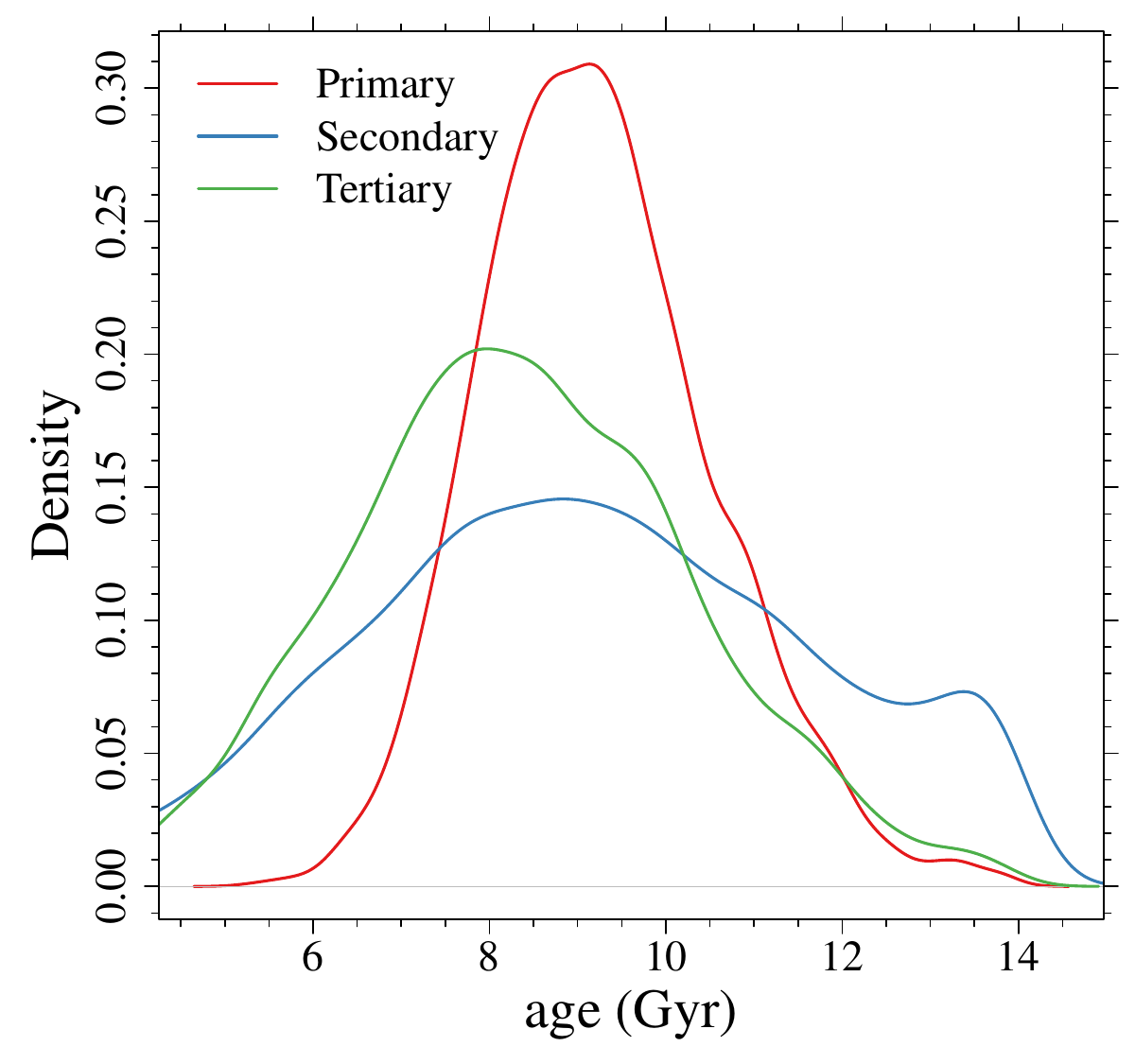}
        \caption{Fit of the  TIC 650024463 system. {\it Left}: 2D kernel density estimator of the posterior distribution of the system parameters in the $Y$ vs age plane. {\it Middle}: Best fitting isochrones in the mass vs radius plane. The points indicate the derived system constraints, while solid red and blue lines correspond to S1 and S2, respectively.   
                {\it Right}: Kernel density estimators of the age of the three stars from independent fits.
        }
        \label{fig:trip1}
\end{figure*}

As often happens with binary systems \citep[e.g.][]{TZFor, Valle2023a}, the posterior distribution of the system parameters exhibits a multimodal density. This is evident in the left panel of Fig.~\ref{fig:trip1}, which shows the probability density in the initial helium abundance $Y$ versus age plane.  A first solution, S1, peaks at low initial helium abundance $Y \approx 0.26$ and an age of $9.35_{-1.18}^{+1.42}$ Gyr, while a second solution, S2, has a higher initial helium abundance $Y \approx 0.28$ and a slightly younger age of $8.76_{-1.07}^{+1.26}$ Gyr. 
Both solutions provide a good fit of the system, with $\chi^2$ values of 1.65 and 0.78, respectively, with six dofs. The small values of the computed $\chi^2$ are primarily attributable to the relatively large uncertainties associated with the system constraints. Specifically, the 1\% precision in the stellar masses and radii not only contributes to a better agreement with the proposed system solution but also affords the algorithm greater flexibility to adjust the fitted masses and radii, improving the agreement with the temperatures and metallicities constraints, thereby enhancing the overall goodness-of-fit. Detailed fit results are collected in Table~\ref{tab:triple1}. The central panel in Fig.~\ref{fig:trip1} shows that the isochrones resulting from the fit are quite similar. 
As expected from theoretical investigations \citep[e.g.][]{binary} the age of the system is mainly set by the more evolved primary star. This is shown in the right panel in Fig.~\ref{fig:trip1} where the kernel density estimators of the age of the stars, from independent fit neglecting the coevality constraints, are shown. The density of the primary star is clearly less dispersed than the other two distributions, leading to an estimated age for the single primary star of $9.15^{+1.40}_{-1.19}$ Gyr. 

The posterior density in Fig.~\ref{fig:trip1} demonstrates the well-known degeneracy between age and initial chemical composition, a common challenge in MS stars age calibration. The probability density gradually decreases away from the peaks, indicating that multiple combinations of age and initial chemical composition can yield nearly equivalent fits for this system.   
Assuming coevality, the averaged age of the system, $9.0_{-1.1}^{+1.4}$ Gyr, is robust because the constraints provided by the three stars with clearly different masses narrow the age range compatible with all of them. The fairly large variability in other parameters, such as the initial helium abundance, only slightly affects the global age. This aligns with the findings of \citet{Valle2024dydz} in binary systems. 
As expected the result of the fit agrees with the underlying astrophysical assumption adopted in \citet{Rappaport2024}, who obtained the system parameters relying on a PARSEC isochrone with age $10.0_{-0.7}^{+0.9}$ Gyr. It is relevant to note that the 14\% precision of our age estimates was obtained with mass constraints at 2\% level. Adopting the same methods, we obtained more precise estimates for binary systems, such as TZ For (6\%) or CPD-54 810 \citep[4\%;][]{TZFor,Valle2023b}, but they required masses and radii observational uncertainties well below 1\%.

\subsection{TIC 323486857}

Constraints for the TIC 323486857, derived from \citet{Rappaport2024}, are presented in Table~\ref{tab:obs-triple2}. As for the previous system, the errors in effective temperature and [Fe/H] were fixed at 100 K and 0.1 dex, respectively.

\begin{table}[ht]
        \centering
        \caption{Constraints for the TIC 323486857 system.} \label{tab:obs-triple2}
        \begin{tabular}{lccc}
                \hline\hline
                $T_{\rm eff}$ (K) & $6550\pm100$        & $6434\pm 100$ & $5068\pm100$ \\
                $M ~ (M_{\sun})$ & $1.408\pm 0.036$     & $ 1.223 \pm 0.038$ &$1.588 \pm 0.040$\\
                $R ~ (R_{\sun})$ & $1.873 \pm 0.07$     & $1.344 \pm 0.08$      & $7.54 \pm 0.58$ \\
                $\rm [Fe/H]$ & $0.01 \pm 0.1$ &&\\
                \hline
        \end{tabular}
\end{table}

\begin{table}[ht]
        \centering
        \caption{Fit of the TIC 323486857 system.} \label{tab:triple2}
        \begin{tabular}{lccc}
                \hline\hline
                        &  S1 & S2 & S3\\
                \hline \\ [-1.8ex]
                $Y$ & $0.269_{-0.008}^{+0.021}$ & $0.285_{-0.021}^{+0.015}$ & $0.263_{-0.003}^{+0.023}$\\ [2pt]
                $Z$ & $0.0146_{-0.0017}^{+0.0022}$ & $0.0164_{-0.0023}^{+0.0025}$ & $0.0138_{-0.0020}^{+0.0019}$\\ [2pt]
                $\beta$ & $0.02\pm0.02$ & $0.28_{-0.02}^{+0.00}$ & \begin{tabular}{c}$0.04_{-0.03}^{+0.08}$\\ $0.13_{-0.10}^{+0.09}$\\ $0.26_{-0.02}^{+0.02}$
                        \end{tabular} \\[2pt]
                $M_{cc,1} ~ (M_{\sun})$ & $0.103_{-0.006}^{+0.008}$ & $0.147_{-0.006}^{+0.004}$ & $0.102_{-0.010}^{+0.017}$ \\[2pt]
                $M_{cc,2} ~ (M_{\sun})$ & $0.021_{-0.010}^{+0.015}$ & $0.111_{-0.007}^{+0.004}$ &  $0.067_{-0.031}^{+0.029}$ \\[2pt]
                Age (Gyr) & $2.23_{-0.15}^{+0.16}$ & $2.39_{-0.17}^{+0.19}$ & $2.33_{-0.16}^{+0.18}$\\ [2pt]
                \hline
                $T_{\rm eff,1}$ (K) & 6470(50) &   6480(54) & 6476(51)  \\ 
                ${\rm [Fe/H]}_1$ & -0.08(7) &   -0.03(6) & 0.02(7)  \\ 
                $M_1 ~ (M_{\sun})$ & 1.40(2) &  1.40(2) & 1.40(2)  \\ 
                $R_1 ~ (R_{\sun})$ & 1.90(5) &   1.92(6) & 1.90(5)  \\ 
                $T_{\rm eff,2}$ (K) & 6398(53) &   6384(56) & 6412(52)  \\ 
                ${\rm [Fe/H]}_2$ & -0.06(7) &   -0.01(7) & 0.02(7)  \\ 
            $M_2 ~ (M_{\sun})$ & 1.21(2) &   1.22(3) & 1.21(2)  \\ 
                $R_2 ~ (R_{\sun})$ & 1.32(3) &   1.36(4)  & 1.32(3) \\ 
                $T_{\rm eff,3}$ (K) & 4894(37) &   4905(40) & 4906(40) \\ 
                ${\rm [Fe/H]}_3$ & 0.05(7) &   0.11(6)  & 0.02(7)\\ 
                $M_3 ~ (M_{\sun})$ & 1.57(3) &   1.57(3) & 1.58(3)  \\ 
                $R_3 ~ (R_{\sun})$ & 6.63(39) &   6.46(36)  & 7.01(34) \\ 
                \hline
                $\chi^2$ & 8.27 &   8.79 & 1.93  \\ 
                $\chi^2_{1,2}$ & 2.44 &   1.55 & 1.08 \\ 
                \hline
        \end{tabular}
    \tablefoot{S1 and S2 are obtained by imposing a common $\beta$ value for all the stars and allowing microscopic diffusion to modify the surface [Fe/H] and adopting 100 K as uncertainty in the tertiary star $T_{\rm eff}$. S3 allows independent $\beta$ values and neglects the effect of microscopic diffusion on surface [Fe/H] and adopts 600 K as uncertainty in the tertiary star $T_{\rm eff}$.}
\end{table}

\begin{figure*}
        \centering
        \includegraphics[height=7.5cm,angle=0]{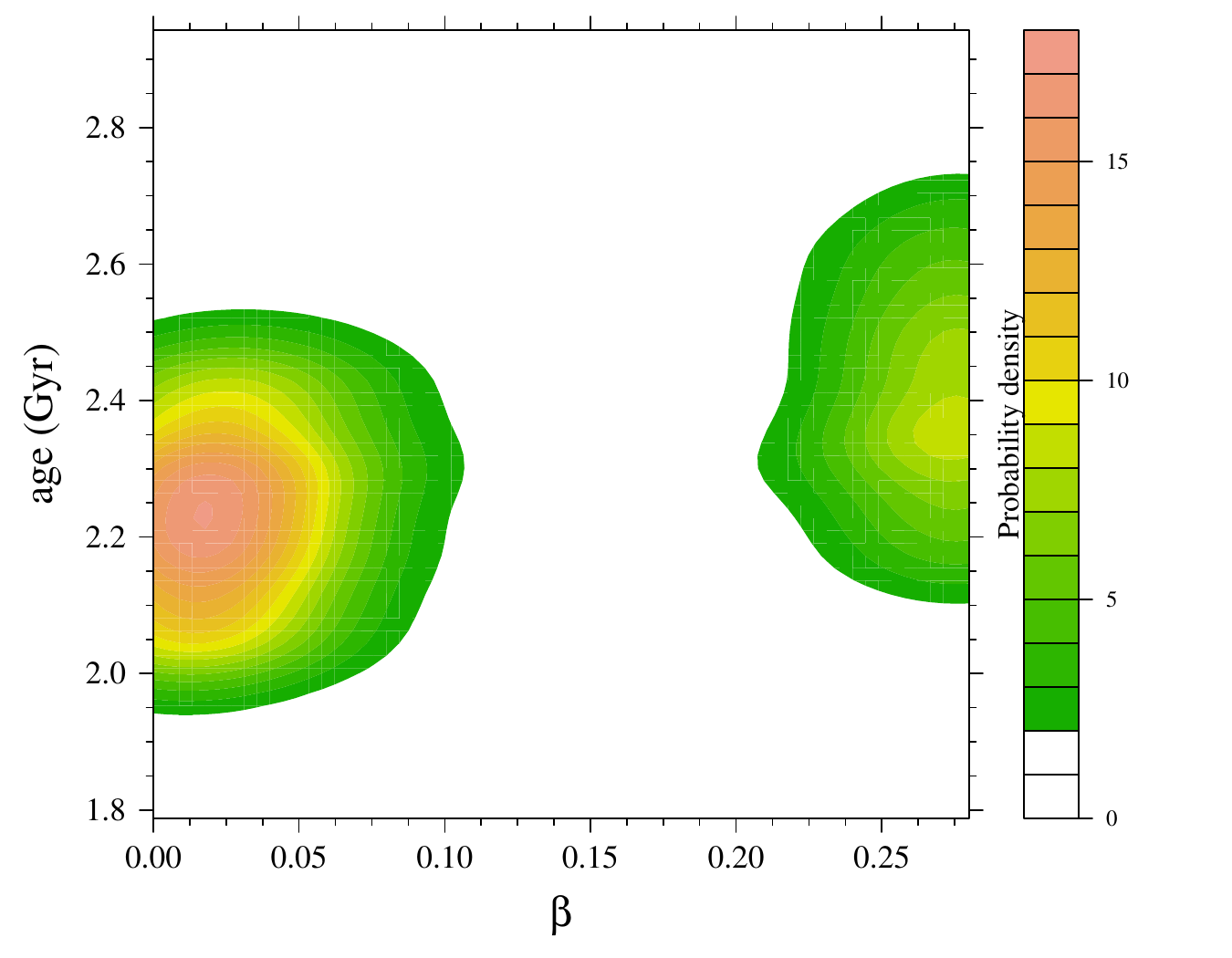}
        \includegraphics[width=8.cm,angle=0]{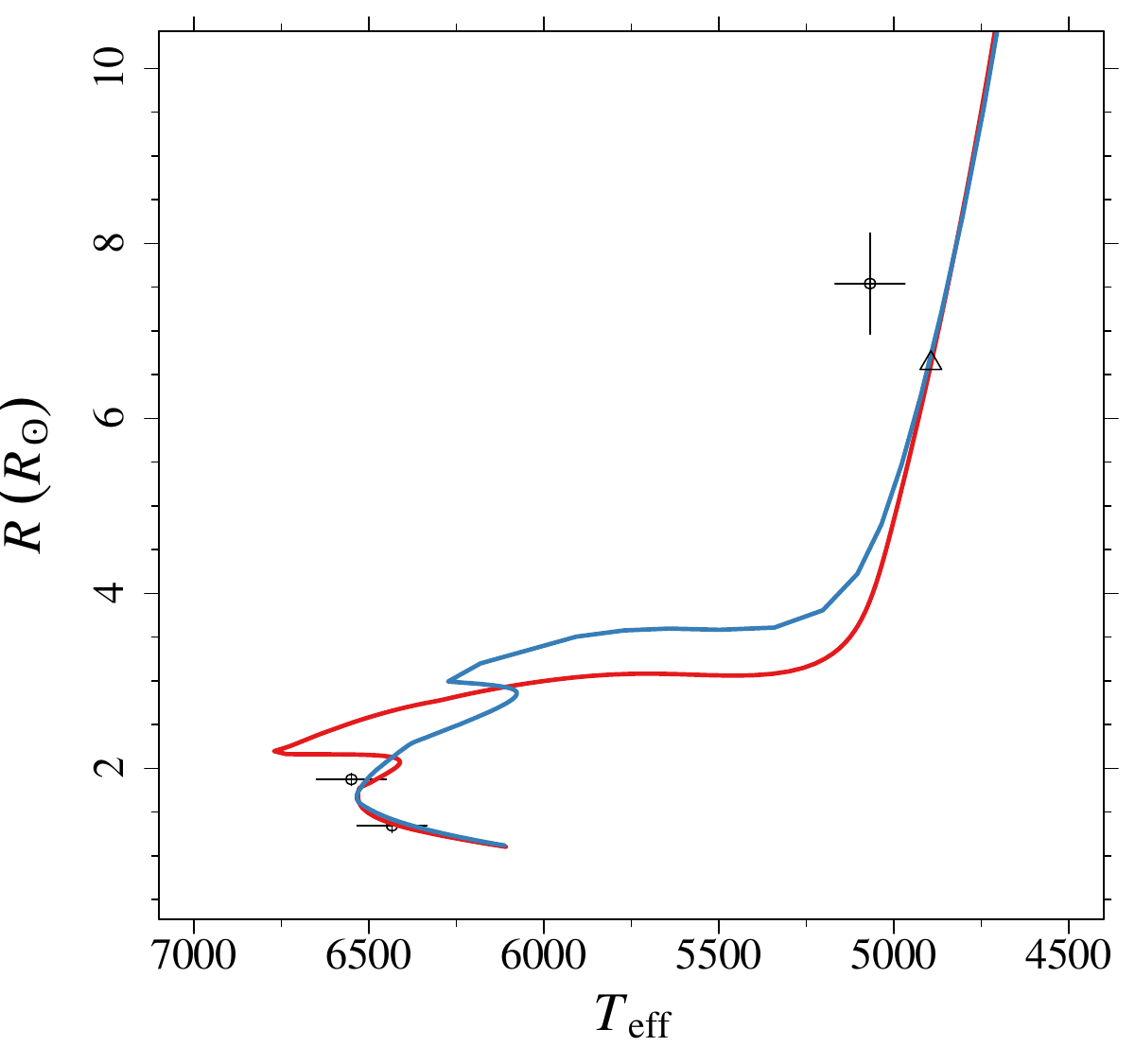}
        \caption{Fit of the  TIC 323486857 system assuming the same core overshooting efficiency for all stars. {\it Left}: 2D kernel density estimator of the posterior distribution of the system parameters in the age vs $\beta$ plane. {\it Right}: Best fitting isochrone in the radius vs effective temperature plane. The open dots indicate the derived system constraints, while the solid red and blue lines correspond to S1 and S2, respectively. The triangle identifies the position of the fitted tertiary star for S1.  
        }
        \label{fig:trip2}
\end{figure*}

\begin{figure}
        \centering
        \includegraphics[height=7.5cm,angle=0]{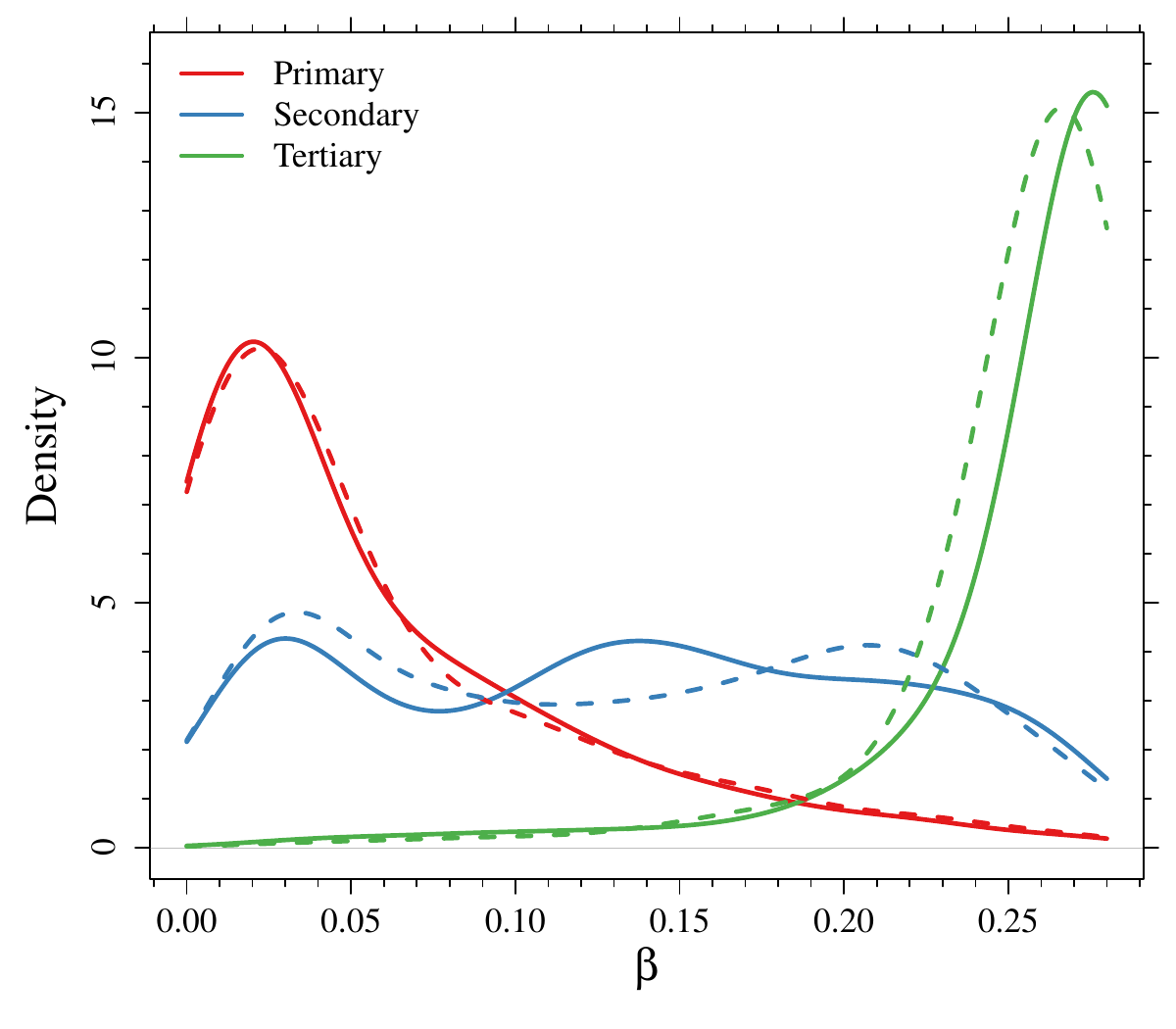}
        \caption{Kernel density estimators of the individual $\beta$ parameters for the three stars in the TIC 323486857 system.  Solid lines represent the fit that adopts the grid with variable surface [Fe/H] and a 100 K uncertainty in the tertiary effective temperature. Dashed lines were obtained by fixing the track grid surface [Fe/H] to the initial values and adopting 600 K as the uncertainty in the tertiary effective temperature. 
        }
        \label{fig:trip2-allov}
\end{figure}

As anticipated, the solution for this system exhibits some limitations and requires supplementary assumptions. As a first fit attempt we tried to constrain the system to have a common core overshooting efficiency for all the stars. This assumption will be further challenged by performing different fits relaxing this as well as other constraints.  

As for the first system, the fit assuming common convective core overshooting efficiency has a bimodal posterior density, but the two solution islands for TIC 323486857 are well separated (left panel in Fig.~\ref{fig:trip2}).
Two solutions are found: S1 at a low $\beta =0.02 \pm 0.02$, and S2 at the edge of the grid, $\beta = 0.28$. The first solution is significantly more probable, and the second solution should be viewed with caution. This is because a broader range of allowed overshooting values in the grid would likely have resulted in a different optimal solution.
Table~\ref{tab:triple2} summarises the full fit parameters, including the convective core masses $M_{cc}$ for the primary and secondary stars (the tertiary has already reached the RGB evolutionary phase, where the convective core disappears).
Both solutions yield high $\chi^2$ values of around 8.5 (with five dofs), largely due to a poor fit for the tertiary star. With a precision of about 4\% in the stellar masses and radii a better agreement is expected, as discussed for the first system. Focusing solely on the inner binary system improves the goodness-of-fit, with $\chi^2_{1,2}$ values around 2. Overall, we get a system age of $2.29_{-0.16}^{+0.17}$ Gyr.
The right panel of Fig.~\ref{fig:trip2} highlights the problematic fit for the tertiary. Its fitted position on the isochrone deviates significantly in effective temperature, with the observational constraint being nearly 150 K hotter than the model prediction. This discrepancy could stem from an inaccurate representation of effective temperature in the RGB phase, potentially due to a not perfect treatment of the superadiabatic convection efficiency \citep{Trampedach2014, Magic2015}.
Notably, the PARSEC isochrones used by \citet{Rappaport2024} closely matches the effective temperature of our isochrones in the RGB phase, suggesting that the discrepancy with this constraint of the system is not a peculiarity of our models.
Alternatively, systematic errors in the system derived parameters  could also explain the discrepancy. However, definitive confirmation requires spectroscopic data to constrain the system's metallicity and effective temperature.
 
An interesting difference exists between our fit and the derived radius of the tertiary star under both solutions. This difference can be attributed to the choice of stellar models used for the observational characterisation of the system by \citet{Rappaport2024}. Specifically, the PARSEC isochrones utilise a different overshooting scheme compared to our models. 
In fact the dependence of the core overshooting parameter on the stellar mass is still debated and got relevant attention in the literature, with opposite claims \citep[see][for a review]{Anders2023}. 
PARSEC isochrones adopt an increasing core overshooting parameter with stellar mass, starting at 1.0 $M_{\sun}$ and reaching a maximum value at 1.5 $M_{\sun}$. 
Moreover, PARSEC parametrisation of the overshooting mechanism is different than our, their maximum values corresponding roughly to $\beta = 0.25$ in our implementation \citep{Bressan2012}.
This choice in the core overshooting efficiency impacts the predicted radius of the tertiary star, leading to a larger value in the \citet{Rappaport2024} system solution compared to ours, which assumes a constant core overshooting efficiency across all stellar masses.   

To further investigate this point, we performed a fit where the algorithm was allowed to choose independent $\beta$ values for each star, with the only constraint being a common initial chemical composition and age. Figure~\ref{fig:trip2-allov} shows that  the solution for the tertiary star peaks at the edge of the grid, while the primary star is well fitted with a low $\beta \approx 0.02$. The poorly evolved secondary star does not provide significant constraints on the overshooting parameter. As in the previous analysis, the $\beta$ for the tertiary stars should be viewed with caution. This is because a better solution may exist at higher overshooting values beyond the range explored in the grid.
The system age estimate of $2.20_{-0.14}^{+0.19}$ Gyr closely matches that obtained in the common overshooting scenario. However, the goodness-of-fit for this system remains questionable, $\chi^2 = 6.1$ (with three dofs) mainly because of the effective temperature tertiary contribution ($\chi^2_{1,2} = 1.8$).  
While the high $\beta$ value for the tertiary star  aligns with what is assumed in the PARSEC isochrones, it is surprisingly high for a 1.6 $M_{\sun}$ star. For comparison with the literature, adopting the same methodology for binary systems,  a maximum estimated overshooting parameter of $\beta \approx 0.17$ was obtained for the more massive TZ For system \citep{TZFor}.  Independently, \citet{Claret2016} suggested that for a star of mass 1.6 $M_{\sun}$ a value of $\beta \approx 0.1$ is expected. 

Overall, none of the analysed fits provide satisfactory agreement with the constraints from \citet{Rappaport2024}. Furthermore, they are potentially affected by several limitations.
First, a poor fit of the tertiary effective temperature, as previously discussed, may significantly influence the results. The discrepancy between theoretical models and the observed temperature can distort the fit, forcing the algorithm to stretch the solution towards the grid edge to accommodate this constraint.
Second, the assumption of a common metallicity for all three stars, derived from PARSEC isochrones, may not be accurate because the stars are in different evolutionary phases. As extensively discussed in \citet{Valle2023a}, the tertiary star's surface metallicity is expected to be close to its initial value. However, the primary and secondary stars, still on the MS, may experience changes in surface metallicity due to competing mixing processes. The adopted grid includes microscopic diffusion, and its effects are evident in the spread of S1 and S2 best-fit [Fe/H] values in Table~\ref{tab:triple2}, exceeding 0.1 dex.

Given the limitations in both the isochrone-derived metallicity and the theoretical predictions of surface [Fe/H] for stars with thin convective envelopes \citep[e.g.][]{Moedas2022}, we performed a supplementary fit to assess the robustness of our results. This fit incorporates a larger 600 K uncertainty in the tertiary effective temperature and fixes the grid surface [Fe/H] for all stellar tracks to the initial value. Therefore, this grid still includes the effect of microscopic diffusion on the evolutionary timescale but neglects its impact on the surface chemical abundances.
The results of this fit, adopting independent overshooting for the three stars, are presented in Table~\ref{tab:triple2} under the S3 scenario. The posterior densities of the convective core overshooting parameters are shown in Fig.~\ref{fig:trip2-allov}, demonstrating agreement with the results discussed previously. The age of the system $2.33^{+0.18}_{-0.16}$ Gyr is in excellent agreement with that determined in the previous scenario.
This analysis suggest a limited impact of alternative  decisions on the recovery of the main system parameters.

Moreover, unlike previous solutions, scenario S3 provides a value for the tertiary $\beta$ parameter (0.26) within the grid, supporting its reliability. Overall, scenario S3 exhibits a satisfactory fit with a $\chi^2$ value of 1.9, partially attributed to the relaxed constraint on the tertiary $T_{\rm eff}$, but also to a better fit for the tertiary radius.
However, since the models are not nested, a formal significance test comparing the likelihoods of different scenarios is not theoretically justified \citep[see e.g.][]{simar, Feigelson2012}.

Encouragingly, all our age estimates and that adopted by \citet{Rappaport2024} in their system solution, $2.37_{-0.20}^{+0.07}$ Gyr, agree very well. This consistency suggests that the system age is a robust determination, even when using different stellar models in the fitting process, provided a sufficiently large parameter grid is employed. 
However, this finding may not apply to systems with stars in different evolutionary phases. In this specific case, the age is mainly dictated by the inner binary system, in particular by the primary star, which contributes with more weight than the secondary to the age estimate. An analysis restricted to the inner binary system shows an age of $2.1_{-0.24}^{+0.33}$ Gyr, close to the estimate obtained when considering also the tertiary star constraint. While the contribution of the tertiary star to fixing the system age is not determinant, the comparison of the error range of this fit with that reported above highlights its importance in getting a precise estimate.

\section{Conclusions}\label{sec:conclusions}

Leveraging recently released data on triply eclipsing triple systems \citep{Rappaport2024}, we investigated the possibility of adopting them to constrain stellar model parameters. 
In particular, we focused on evaluating the influence of the astrophysical assumptions employed  for the observational characterisation of the system by \citet{Rappaport2024}.

We utilised dense grids of pre-computed stellar models to fit the data for the triply eclipsing systems with a modified version of the SCEPtER pipeline adapted for triple systems. For the first system, TIC 650024463, which comprises three low-mass MS stars, we achieved an excellent agreement with observational data at $9.0_{-1.1}^{+1.4}$ Gyr. As commonly observed in binary system fitting, the algorithm identified a multimodal posterior density for the fit parameters, suggesting the presence of two plausible solutions with differing initial helium abundances. 
A good agreement with the system constraints derived by \citet{Rappaport2024} was expected because the difference between the theoretical stellar models adopted by \citet{Rappaport2024} and the present ones used for the system fit are minor for low-mass MS stars   \citep{incertezze1, Stancliffe2015}. 

Fitting the second system, TIC 323486857, proved more challenging. This system comprises stars with masses ranging from about
1.2 $M_{\sun}$ to  1.6 $M_{\sun}$; the tertiary and most massive star is in the RGB phase. 
We tested three fitting scenarios: (i) assuming a constant efficiency for convective core overshooting across all stellar masses, (ii) allowing independent overshooting efficiencies for each star, and (iii) assuming a different effective temperature precision for the tertiary star and a different efficiency for the mixing processes regarding the evolution of the surface chemical composition, keeping the independent overshooting efficiencies for each star.

In the first scenario, with a constant overshooting efficiency, the fit for the tertiary star's parameters, particularly its effective temperature and radius, displayed significant discrepancies with the observed data. Notably, our derived effective temperature aligns with the astrophysical models used in \citet{Rappaport2024} to determine the system parameters. This suggests that the discrepancy is not attributable to the underlying difference in the RGB temperature scale between our models and PARSEC isochrones.  
As a matter of fact, systematic differences in the effective temperature between models and observations in the RGB phase are reported in literature \citep[see][and references therein for a recent analysis]{Valle2024age}.
Further investigation on this topic will require spectroscopic data for the system. 
The discrepancy in the tertiary star's radius is particularly intriguing, as it likely arises from differing assumptions regarding the convective core overshooting efficiency between our models and those employed by \citet{Rappaport2024}.  Their analysis utilises PARSEC isochrones, which incorporate an increasing convective core overshooting efficiency with stellar mass and imply a larger radius for the RGB star.

When we allowed the overshooting parameter to vary freely, as in second and third scenarios, we achieved a close match to the assumptions of the PARSEC models. 
A key difference between these scenarios is that the second includes the effect of microscopic diffusion on the surface metallicity, while the third one does not. The third scenario offers the most robust system parameter determination given the fact that the metallicity constraint is derived from an isochrone and not from spectroscopic individual determinations.  
PARSEC models assume a convective core overshooting parameter of $\beta \approx 0.25$ for the tertiary star, and our analysis of the third scenario yielded  $\beta \approx 0.26$.
While this value aligns well with the PARSEC model assumptions, it is notably higher than what is typically reported in the literature for a star of 1.6 $M_{\sun}$. For instance, \citet{Claret2016} propose a value of $\beta \approx 0.1$. Furthermore, the third scenario, which mimics the assumptions of PARSEC isochrones as closely as possible, provides a clean fit of the system, significantly improving the tertiary radius fit.  
This finding suggests that the underlying astrophysical assumptions in the \citet{Rappaport2024} derivation of the system constraints exert a dominant influence on the fitted $\beta$ parameter. 
While analysing other triply eclipsing triple systems would be extremely valuable for assessing the generality of this result, the substantial computational cost of fitting these systems limited our investigation. Further independent research on this topic could provide deeper insights into the impact of astrophysical assumptions on the fitting of triply eclipsing triple systems.
Encouragingly, despite the relevant differences in the adopted stellar models, our most robust age determination, $2.33_{-0.16}^{+0.18}$ Gyr, agrees well with that from the models in  \citet{Rappaport2024}.

In conclusion, our independent analysis suggests a minimal influence of the astrophysical models used by \citet{Rappaport2024} to obtain stellar system data on the derived parameters for MS stars, particularly those of low mass.  However, the suitability of these systems for testing stellar evolution in more massive and evolved stars is still an open question.
Encouragingly, the system ages derived here show good agreement with those used in the \citet{Rappaport2024} system parameter determination, indicating a relatively robust age determination.  While this conclusion may not hold for systems with stars in different evolutionary stages, it appears that triply eclipsing triple system data, obtained without RV measurements,  hold promise to constrain stellar ages, provided a sufficiently extensive parameter grid is employed during the fitting process.

\begin{acknowledgements}
We thank the referees for the insightful comments and constructive suggestions.
G.V., P.G.P.M. and S.D. acknowledge INFN (Iniziativa specifica TAsP) and support from PRIN MIUR2022 Progetto "CHRONOS" (PI: S. Cassisi) finanziato dall'Unione Europea - Next Generation EU.
\end{acknowledgements}

\bibliographystyle{aa}
\bibliography{allbib}

\begin{thebibliography}{36}
\expandafter\ifx\csname natexlab\endcsname\relax\def\natexlab#1{#1}\fi

\bibitem[{{Alonso} {et~al.}(2015){Alonso}, {Deeg}, {Hoyer}, {Lodieu}, {Palle},
  \& {Sanchis-Ojeda}}]{Alonso2015}
{Alonso}, R., {Deeg}, H.~J., {Hoyer}, S., {et~al.} 2015, \aap, 584, L8

\bibitem[{{Anders} \& {Pedersen}(2023)}]{Anders2023}
{Anders}, E.~H. \& {Pedersen}, M.~G. 2023, Galaxies, 11, 56

\bibitem[{{Asplund} {et~al.}(2009){Asplund}, {Grevesse}, {Sauval}, \&
  {Scott}}]{AGSS09}
{Asplund}, M., {Grevesse}, N., {Sauval}, A.~J., \& {Scott}, P. 2009, \araa, 47,
  481

\bibitem[{{Borkovits} {et~al.}(2022){Borkovits}, {Mitnyan}, {Rappaport},
  {Pribulla}, {Powell}, {Kostov}, {B{\'\i}r{\'o}}, {Cs{\'a}nyi}, {Garai},
  {Gary}, {Kaye}, {Kom{\v{z}}{\'\i}k}, {Terentev}, {Omohundro}, {Gagliano},
  {Jacobs}, {Kristiansen}, {LaCourse}, {Schwengeler}, {Czavalinga}, {Seli},
  {Huang}, {P{\'a}l}, {Vanderburg}, {Rodriguez}, \& {Stevens}}]{Borkovits2022}
{Borkovits}, T., {Mitnyan}, T., {Rappaport}, S.~A., {et~al.} 2022, \mnras, 510,
  1352

\bibitem[{{Borkovits} {et~al.}(2019){Borkovits}, {Rappaport}, {Kaye},
  {Isaacson}, {Vanderburg}, {Howard}, {Kristiansen}, {Omohundro},
  {Schwengeler}, {Terentev}, {Shporer}, {Relles}, {Villanueva}, {Tan},
  {Col{\'o}n}, {Blex}, {Haas}, {Cochran}, \& {Endl}}]{Borkovits2019}
{Borkovits}, T., {Rappaport}, S., {Kaye}, T., {et~al.} 2019, \mnras, 483, 1934

\bibitem[{{Bressan} {et~al.}(2012){Bressan}, {Marigo}, {Girardi}, {Salasnich},
  {Dal Cero}, {Rubele}, \& {Nanni}}]{Bressan2012}
{Bressan}, A., {Marigo}, P., {Girardi}, L., {et~al.} 2012, \mnras, 427, 127

\bibitem[{{Carter} {et~al.}(2011){Carter}, {Fabrycky}, {Ragozzine}, {Holman},
  {Quinn}, {Latham}, {Buchhave}, {Van Cleve}, {Cochran}, {Cote}, {Endl},
  {Ford}, {Haas}, {Jenkins}, {Koch}, {Li}, {Lissauer}, {MacQueen}, {Middour},
  {Orosz}, {Rowe}, {Steffen}, \& {Welsh}}]{Carter2011}
{Carter}, J.~A., {Fabrycky}, D.~C., {Ragozzine}, D., {et~al.} 2011, Science,
  331, 562

\bibitem[{{Chaboyer} {et~al.}(2001){Chaboyer}, {Fenton}, {Nelan}, {Patnaude},
  \& {Simon}}]{Chaboyer2001}
{Chaboyer}, B., {Fenton}, W.~H., {Nelan}, J.~E., {Patnaude}, D.~J., \& {Simon},
  F.~E. 2001, \apj, 562, 521

\bibitem[{{Claret} \& {Torres}(2016)}]{Claret2016}
{Claret}, A. \& {Torres}, G. 2016, \aap, 592, A15

\bibitem[{{Degl'Innocenti} {et~al.}(2008){Degl'Innocenti}, {Prada Moroni},
  {Marconi}, \& {Ruoppo}}]{scilla2008}
{Degl'Innocenti}, S., {Prada Moroni}, P.~G., {Marconi}, M., \& {Ruoppo}, A.
  2008, \apss, 316, 25

\bibitem[{{Dell'Omodarme} {et~al.}(2012){Dell'Omodarme}, {Valle},
  {Degl'Innocenti}, \& {Prada Moroni}}]{database2012}
{Dell'Omodarme}, M., {Valle}, G., {Degl'Innocenti}, S., \& {Prada Moroni},
  P.~G. 2012, A\&A, 540, A26

\bibitem[{{Derekas} {et~al.}(2011){Derekas}, {Kiss}, {Borkovits}, {Huber},
  {Lehmann}, {Southworth}, {Bedding}, {Balam}, {Hartmann}, {Hrudkova},
  {Ireland}, {Kov{\'a}cs}, {Mez{\H{o}}}, {Mo{\'o}r}, {Niemczura}, {Sarty},
  {Szab{\'o}}, {Szab{\'o}}, {Telting}, {Tkachenko}, {Uytterhoeven},
  {Benk{\H{o}}}, {Bryson}, {Maestro}, {Simon}, {Stello}, {Schaefer}, {Aerts},
  {ten Brummelaar}, {De Cat}, {McAlister}, {Maceroni}, {M{\'e}rand}, {Still},
  {Sturmann}, {Sturmann}, {Turner}, {Tuthill}, {Christensen-Dalsgaard},
  {Gilliland}, {Kjeldsen}, {Quintana}, {Tenenbaum}, \& {Twicken}}]{Derekas2011}
{Derekas}, A., {Kiss}, L.~L., {Borkovits}, T., {et~al.} 2011, Science, 332, 216

\bibitem[{{Feiden} {et~al.}(2011){Feiden}, {Chaboyer}, \&
  {Dotter}}]{Feiden2011}
{Feiden}, G.~A., {Chaboyer}, B., \& {Dotter}, A. 2011, \apjl, 740, L25

\bibitem[{Feigelson \& Babu(2012)}]{Feigelson2012}
Feigelson, E.~D. \& Babu, G.~J. 2012, Modern Statistical Methods for Astronomy
  with R applications (Cambridge University Press)

\bibitem[{H{\"a}rdle \& Simar(2012)}]{simar}
H{\"a}rdle, W.~K. \& Simar, L. 2012, Applied Multivariate Statistical Analysis
  (Springer)

\bibitem[{{He{\l}miniak} {et~al.}(2021){He{\l}miniak}, {Moharana}, {Pawar},
  {Ukita}, {Sybilski}, {Espinoza}, {Kambe}, {Ratajczak}, {Jord{\'a}n},
  {Maehara}, {Brahm}, {Koz{\l}owski}, \& {Konacki}}]{Helminiak2021}
{He{\l}miniak}, K.~G., {Moharana}, A., {Pawar}, T., {et~al.} 2021, \mnras, 508,
  5687

\bibitem[{{Magic} {et~al.}(2015){Magic}, {Weiss}, \& {Asplund}}]{Magic2015}
{Magic}, Z., {Weiss}, A., \& {Asplund}, M. 2015, \aap, 573, A89

\bibitem[{{Miller} {et~al.}(2020){Miller}, {Maxted}, \& {Smalley}}]{Miller2020}
{Miller}, N.~J., {Maxted}, P.~F.~L., \& {Smalley}, B. 2020, \mnras, 497, 2899

\bibitem[{{Moedas} {et~al.}(2022){Moedas}, {Deal}, {Bossini}, \&
  {Campilho}}]{Moedas2022}
{Moedas}, N., {Deal}, M., {Bossini}, D., \& {Campilho}, B. 2022, \aap, 666, A43

\bibitem[{{Planck Collaboration} {et~al.}(2020){Planck Collaboration},
  {Aghanim}, {Akrami}, {Ashdown}, {Aumont}, {Baccigalupi}, {Ballardini},
  {Banday}, {Barreiro}, {Bartolo}, {Basak}, {Battye}, {Benabed}, {Bernard},
  {Bersanelli}, {Bielewicz}, {Bock}, {Bond}, {Borrill}, {Bouchet}, {Boulanger},
  {Bucher}, {Burigana}, {Butler}, {Calabrese}, {Cardoso}, {Carron},
  {Challinor}, {Chiang}, {Chluba}, {Colombo}, {Combet}, {Contreras}, {Crill},
  {Cuttaia}, {de Bernardis}, {de Zotti}, {Delabrouille}, {Delouis}, {Di
  Valentino}, {Diego}, {Dor{\'e}}, {Douspis}, {Ducout}, {Dupac}, {Dusini},
  {Efstathiou}, {Elsner}, {En{\ss}lin}, {Eriksen}, {Fantaye}, {Farhang},
  {Fergusson}, {Fernandez-Cobos}, {Finelli}, {Forastieri}, {Frailis},
  {Fraisse}, {Franceschi}, {Frolov}, {Galeotta}, {Galli}, {Ganga},
  {G{\'e}nova-Santos}, {Gerbino}, {Ghosh}, {Gonz{\'a}lez-Nuevo}, {G{\'o}rski},
  {Gratton}, {Gruppuso}, {Gudmundsson}, {Hamann}, {Handley}, {Hansen},
  {Herranz}, {Hildebrandt}, {Hivon}, {Huang}, {Jaffe}, {Jones}, {Karakci},
  {Keih{\"a}nen}, {Keskitalo}, {Kiiveri}, {Kim}, {Kisner}, {Knox},
  {Krachmalnicoff}, {Kunz}, {Kurki-Suonio}, {Lagache}, {Lamarre}, {Lasenby},
  {Lattanzi}, {Lawrence}, {Le Jeune}, {Lemos}, {Lesgourgues}, {Levrier},
  {Lewis}, {Liguori}, {Lilje}, {Lilley}, {Lindholm}, {L{\'o}pez-Caniego},
  {Lubin}, {Ma}, {Mac{\'\i}as-P{\'e}rez}, {Maggio}, {Maino}, {Mandolesi},
  {Mangilli}, {Marcos-Caballero}, {Maris}, {Martin}, {Martinelli},
  {Mart{\'\i}nez-Gonz{\'a}lez}, {Matarrese}, {Mauri}, {McEwen}, {Meinhold},
  {Melchiorri}, {Mennella}, {Migliaccio}, {Millea}, {Mitra},
  {Miville-Desch{\^e}nes}, {Molinari}, {Montier}, {Morgante}, {Moss}, {Natoli},
  {N{\o}rgaard-Nielsen}, {Pagano}, {Paoletti}, {Partridge}, {Patanchon},
  {Peiris}, {Perrotta}, {Pettorino}, {Piacentini}, {Polastri}, {Polenta},
  {Puget}, {Rachen}, {Reinecke}, {Remazeilles}, {Renzi}, {Rocha}, {Rosset},
  {Roudier}, {Rubi{\~n}o-Mart{\'\i}n}, {Ruiz-Granados}, {Salvati}, {Sandri},
  {Savelainen}, {Scott}, {Shellard}, {Sirignano}, {Sirri}, {Spencer},
  {Sunyaev}, {Suur-Uski}, {Tauber}, {Tavagnacco}, {Tenti}, {Toffolatti},
  {Tomasi}, {Trombetti}, {Valenziano}, {Valiviita}, {Van Tent}, {Vibert},
  {Vielva}, {Villa}, {Vittorio}, {Wandelt}, {Wehus}, {White}, {White},
  {Zacchei}, \& {Zonca}}]{Planck2020}
{Planck Collaboration}, {Aghanim}, N., {Akrami}, Y., {et~al.} 2020, \aap, 641,
  A6

\bibitem[{{Rappaport} {et~al.}(2023){Rappaport}, {Borkovits}, {Gagliano},
  {Jacobs}, {Tokovinin}, {Mitnyan}, {Kom{\v{z}}{\'\i}k}, {Kostov}, {Powell},
  {Torres}, {Terentev}, {Omohundro}, {Pribulla}, {Vanderburg}, {Kristiansen},
  {Latham}, {Schwengeler}, {LaCourse}, {B{\'\i}r{\'o}}, {Cs{\'a}nyi},
  {Czavalinga}, {Garai}, {P{\'a}l}, {Rodriguez}, \& {Stevens}}]{Rappaport2023}
{Rappaport}, S.~A., {Borkovits}, T., {Gagliano}, R., {et~al.} 2023, \mnras,
  521, 558

\bibitem[{{Rappaport} {et~al.}(2024){Rappaport}, {Borkovits}, {Mitnyan},
  {Gagliano}, {Eisner}, {Jacobs}, {Tokovinin}, {Powell}, {Kostov}, {Omohundro},
  {Kristiansen}, {Jayaraman}, {Terentev}, {Schwengeler}, {LaCourse}, {Garai},
  {Pribulla}, {Maxted}, {B{\'\i}r{\'o}}, {Cs{\'a}nyi}, {P{\'a}l}, \&
  {Vanderburg}}]{Rappaport2024}
{Rappaport}, S.~A., {Borkovits}, T., {Mitnyan}, T., {et~al.} 2024, \aap, 686,
  A27

\bibitem[{{Salaris} \& {Cassisi}(2015)}]{Salaris2015}
{Salaris}, M. \& {Cassisi}, S. 2015, \aap, 577, A60

\bibitem[{{Salaris} {et~al.}(2018){Salaris}, {Cassisi}, {Schiavon}, \&
  {Pietrinferni}}]{Salaris2018}
{Salaris}, M., {Cassisi}, S., {Schiavon}, R.~P., \& {Pietrinferni}, A. 2018,
  \aap, 612, A68

\bibitem[{{Stancliffe} {et~al.}(2015){Stancliffe}, {Fossati}, {Passy}, \&
  {Schneider}}]{Stancliffe2015}
{Stancliffe}, R.~J., {Fossati}, L., {Passy}, J.-C., \& {Schneider}, F.~R.~N.
  2015, \aap, 575, A117

\bibitem[{{Thoul} {et~al.}(1994){Thoul}, {Bahcall}, \& {Loeb}}]{thoul94}
{Thoul}, A.~A., {Bahcall}, J.~N., \& {Loeb}, A. 1994, \apj, 421, 828

\bibitem[{{Trampedach} {et~al.}(2014){Trampedach}, {Stein},
  {Christensen-Dalsgaard}, {Nordlund}, \& {Asplund}}]{Trampedach2014}
{Trampedach}, R., {Stein}, R.~F., {Christensen-Dalsgaard}, J., {Nordlund},
  {\AA}., \& {Asplund}, M. 2014, \mnras, 445, 4366

\bibitem[{{Valle} {et~al.}(2013){Valle}, {Dell'Omodarme}, {Prada Moroni}, \&
  {Degl'Innocenti}}]{incertezze1}
{Valle}, G., {Dell'Omodarme}, M., {Prada Moroni}, P.~G., \& {Degl'Innocenti},
  S. 2013, \aap, 549, A50

\bibitem[{{Valle} {et~al.}(2015){Valle}, {Dell'Omodarme}, {Prada Moroni}, \&
  {Degl'Innocenti}}]{binary}
{Valle}, G., {Dell'Omodarme}, M., {Prada Moroni}, P.~G., \& {Degl'Innocenti},
  S. 2015, \aap, 579, A59

\bibitem[{{Valle} {et~al.}(2017){Valle}, {Dell'Omodarme}, {Prada Moroni}, \&
  {Degl'Innocenti}}]{TZFor}
{Valle}, G., {Dell'Omodarme}, M., {Prada Moroni}, P.~G., \& {Degl'Innocenti},
  S. 2017, \aap, 600, A41

\bibitem[{{Valle} {et~al.}(2023{\natexlab{a}}){Valle}, {Dell'Omodarme}, {Prada
  Moroni}, \& {Degl'Innocenti}}]{Valle2023b}
{Valle}, G., {Dell'Omodarme}, M., {Prada Moroni}, P.~G., \& {Degl'Innocenti},
  S. 2023{\natexlab{a}}, \aap, 673, A133

\bibitem[{{Valle} {et~al.}(2023{\natexlab{b}}){Valle}, {Dell'Omodarme}, {Prada
  Moroni}, \& {Degl'Innocenti}}]{Valle2023a}
{Valle}, G., {Dell'Omodarme}, M., {Prada Moroni}, P.~G., \& {Degl'Innocenti},
  S. 2023{\natexlab{b}}, \aap, 678, A203

\bibitem[{{Valle} {et~al.}(2024{\natexlab{a}}){Valle}, {Dell'Omodarme}, {Prada
  Moroni}, \& {Degl'Innocenti}}]{Valle2024dydz}
{Valle}, G., {Dell'Omodarme}, M., {Prada Moroni}, P.~G., \& {Degl'Innocenti},
  S. 2024{\natexlab{a}}, \aap, 687, A294

\bibitem[{{Valle} {et~al.}(2024{\natexlab{b}}){Valle}, {Dell'Omodarme}, {Prada
  Moroni}, \& {Degl'Innocenti}}]{Valle2024age}
{Valle}, G., {Dell'Omodarme}, M., {Prada Moroni}, P.~G., \& {Degl'Innocenti},
  S. 2024{\natexlab{b}}, \aap, 690, A323

\bibitem[{{Valle} {et~al.}(2021){Valle}, {Dell'Omodarme}, \&
  {Tognelli}}]{goodness2021}
{Valle}, G., {Dell'Omodarme}, M., \& {Tognelli}, E. 2021, \aap, 649, A127

\bibitem[{{Vernazza} {et~al.}(1981){Vernazza}, {Avrett}, \&
  {Loeser}}]{Vernazza1981}
{Vernazza}, J.~E., {Avrett}, E.~H., \& {Loeser}, R. 1981, \apjs, 45, 635

\end{thebibliography}

\end{document}